\documentclass[11pt,a4paper]{article}
\usepackage[hyperref]{acl2021}
\usepackage{times}
\usepackage{latexsym}

\usepackage{microtype}
\usepackage{multirow}
\usepackage{array}
\usepackage{graphicx}
\usepackage{amsmath}

\aclfinalcopy % Uncomment this line for the final submission
\def\aclpaperid{522} %  Enter the acl Paper ID here

\title{Meet The Truth: Leverage Objective Facts and Subjective Views for Interpretable Rumor Detection}

\author{Jiawen Li, Shiwen Ni and Hung-Yu Kao \\
	Intelligent Knowledge Management Lab\\
  Department of Computer Science and Information Engineering
   \\
  National Cheng Kung University \\
  Tainan, Taiwan \\
  \texttt{\{P78073012,P78083033\}@gs.ncku.edu.tw,hykao@mail.ncku.edu.tw} \\}

\date{}

\begin{document}

\maketitle
\begin{abstract}
Most of the current adversarial training is to minimize the maximum risk of input perturbations, which has been proved to be a regularization method to improve the generalization ability of models. However, only the input attack is too singular, and we spread the attack to the weight parameters of neural networks. In this work, we propose a new adversarial training method, DropAttack, which is inspired by the idea of dropout to allow a certain weight parameter of the model to be attacked with a certain probability, and the generalization of neural network is improved by minimizing the adversarial risk of weight attacks. To validate the effectiveness of the proposed method, we used five public datasets in the fields of natural language processing and computer vision for experimental testing. The experimental results show that DropAttack improves generalization performance on all datasets compared to neural networks that do not use DropAttack. In addition, we compare the proposed method with other adversarial training methods and regularization methods, and our method achieves state-of-the-art on all datasets.
\end{abstract}

\section{Introduction}

With the prevalence of social media platforms, rumors have been a serious social problem. Notably, existing rumor detection methods roughly formulate this task as a natural language classification task. The goal of the task is to simply label a given textual claim as rumor or non-rumor. Nevertheless, only a verdict to a suspicious statement is insufficient for people to understand and reason why a claim is a rumor.
For example, Fig.1 is the comparison figure of existing rumor detection methods and a rumor detection method that provides evidence. The claim in Fig. 1 is a half-truth, which is highly deceptive. For such rumors, providing a label only is unconvincing. Thus, we believe that a good rumor detection system should have 2 essential functions including, a rumor identifying function and evidence providing function.

Rumor detection that provides evidence has the following benefits: (1) Improve detection performance. (2) Improve the user experience. (3) Provide a basis for manual review. (4) Improve the accuracy of early rumor detection. (5) Intercept the spread of similar rumors.
%图1
\begin{figure}[t]
	\centering
	\includegraphics[width=1\linewidth]{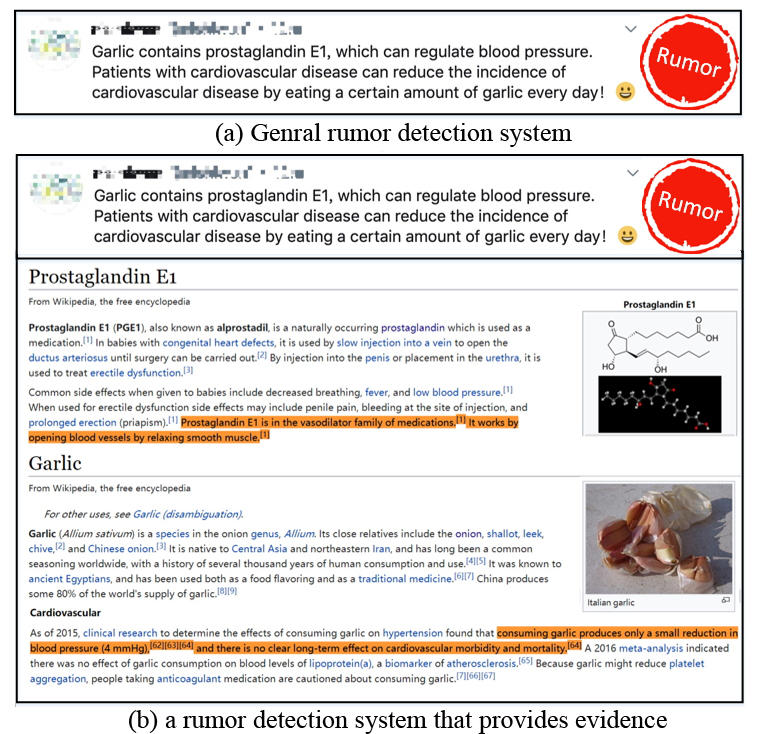}
	\caption{Comparison figure of rumor detection results of two different systems. The orange highlights in Fig.1 (b) are pieces of evidence retrieved from Wikipedia. From those evidence sentences, readers can easily judge if the given claim is a half-truth and clearly understand why that claim is a rumor.}
	\label{fig1}
\end{figure}

Despite having numerous advantages, rumor detection that provides evidence is extremely hard. If none of the labeled evidence information is included in a rumor detection training dataset, the deep learning network is unlikely to generate these textual evidence contents by itself. Unfortunately, the datasets currently used for rumor detection cannot be used as evidence. 

To find out what type of information that can be used as evidence, two different kinds of information, subjective information and objective information, are discussed in this part \citep{merigo2016subjective,zorio2020consumer}. Under the field of rumor detection, subjective information refers to source tweets, comments, etc, while objective information refers to the information on Wikipedia or Baidu Encyclopedia, etc.
Through our comprehensive analysis, we found that subjective information and objective information shows distinct-different characteristics, which are summarized in Table 1. The objective information is consistency and high purity, can be used as evidence, and the subjective information also contains certain clues for debunking rumors.

% 表2
\begin{table}[t]
	\centering
	\caption{Comparison characteristics table of subjective and objective
		information.}
	\label{table}
	\setlength{\tabcolsep}{2pt}
	\begin{tabular}{|p{105pt}|p{105pt}|}
		\hline
		\textbf{Subjectivity infor}& 
		\textbf{Objective infor} \\
		\hline
		easy access	& 
		need crawl \\
		\hline
		extensive	& 
		rare \\
		\hline
		one-sidedness	& 
		comprehensive \\
		\hline
		conflicting	& 
		consistency\\
		\hline
		has noise	& 
		high purity\\
		\hline
	\end{tabular}
	\label{tab1}
\end{table}

To take advantage of both subjective information and objective information, a novel model, LOSIRD, is proposed in this paper. This is notoriously challenging, the difficulties lie in: (1) The model should have a strong retrieval ability. (2) The model should have a Natural Language Inference(NLI) ability. (3) The model needs to be able to process the topology information.

% 图3
\begin{figure*}[t]
	\centering
	\includegraphics[width=0.95\linewidth]{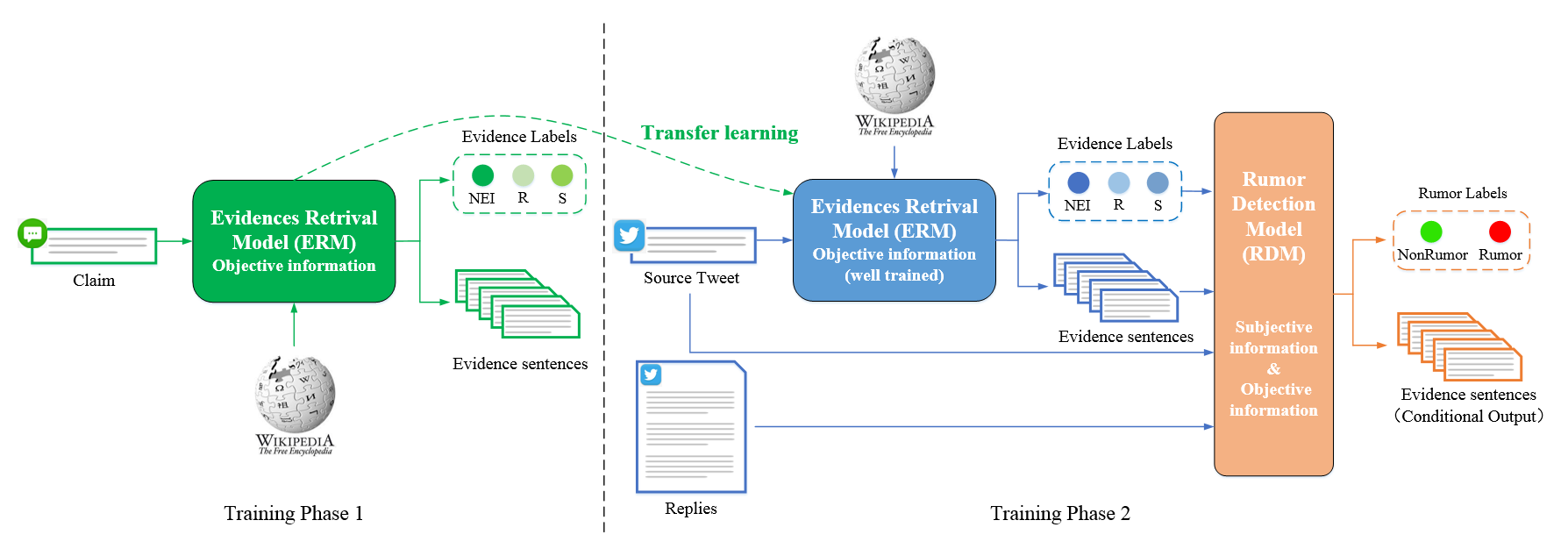}
	\caption{The architecture of our LOSIRD model. The claim and source tweet are essentially the same thing in this paper, “Claim” means the model is training using the Fever dataset, while  “source tweet”  indicates the PHEME datasets are used. }
	\label{fig:f3}
\end{figure*}

Fig.2 shows a high-level view of 
its architecture. This model is 
divided into two modules i.e., ERM (Evidence Retrieval Module) and RDM (Rumor Detection Module).
Inspired by the concept of transfer learning, a two-stage training approach was used for our LOSIRD model. In the first training phase, a widely used fact-checking database was utilized for training the ERM module. In the second training phase, two rumor detection datasets were used to train and evaluate the model. 

The main contribution of this paper is four folds:

\begin{enumerate}
	\item This study, for the first time, arguably proposes a rumor detection model that provides evidence. 
	\item We are the first to propose two novel graph objects to simulate the  propagation lay out of tweets and embedding the relationship of evidence and the claim rumor detection task.
	\item Our LOSIRD achieved the highest detection accuracy and outperformed state-of-the-art models in the rumor detection task.
	\item Our LOSIRD is more generalizable and robust in the early detection of rumor. 
\end{enumerate}

\section{Related work}
\subsection{Evidence Retrieval}
The evidence retrieval task is highly correlated with the rumor detection task. One of the most widely used datasets for evidence retrieval is FEVER\footnote[1]{http://fever.ai/task.html}. 
Majority of researchers handle the fever share task by following the FEVER organizers’ pipeline approach, retrieve and verify the evidence in three steps \citep{hanselowski2018ukp,malon2018team}. \citet{zhou2019gear} formulated claim verification as a graph reasoning task and provides two kinds of attention. \citet{liu2020fine} presented KGAT combining edge kernels and node kernels to better embedding and filtering the evidence. \citet{zhong2020reasoning} constructed two semantic-level topologies to enhance verification performance.  \citet{yoneda2018ucl} employed a four-stage model for the fever share task.

\subsection{Rumor Detection}
The existing rumor detection deep learning methods can be divided into three categories, feature-driven method, content-driven method, and hybrid-driven method.

\textbf{Feature-Driven} approaches, like machine learning methods, rely on a variety of characteristics to identify rumors. \citet{rath2017retweet} proposed a new concept of believability for automatic identification of the users spreading rumors. 

\textbf{Content-Driven} approaches are a kind of method base on natural language processing. Many researchers adopted deep learning models to handle this task \citep{rath2017retweet,ma2016detecting,yu2017convolutional,chen2018call,ma2018rumor}. \citet{monti2019fake} proposed a propagation-based Fake News Detection by GCN. \citet{nguyen2019graph} detected rumor using Multi-modal Social Graph. \citet{sujana2020rumor} proposed a multiloss hierarchical BiLSTM model for fake news detection.

\textbf{Hybrid-Driven} approaches incorporate both feature engineering and text information representation to detect rumors \citep{liu2018early,yang2018ti}. \citet{ruchansky2017csi} proposed a model called CSI for rumor detection, which uses articles and extracts user characteristics to debunk rumors. \citet{lu2020gcan} classified rumor by extracting user's features from their profiles and social interactions. \citet{li2020exploiting} used GraphSEGA to encode the conversation structure. \citet{li2020birds} crawled user-follower information and built a friendly network based on the follow-followers relationship. \citet{castillo2011information} used tweets and re-posts information to detect rumors. \citet{kochkina2018all,li2019rumor} proposed a multi-task learning method to joint training of the main and auxiliary tasks, improving model’s rumor detection the performance. \citet{liu2015real} aggregated common sense and investigative journalism of Twitter users for rumor detection. \citet{ma2017detect} encoded post’s propagation structure for rumor detection. Detect rumors in microblog posts using propagation structure via kernel learning.

\subsection{Comparison}
The highlights of our model include providing evidence, covering two heterogeneous structure graph information, combining both evidence clues and replies information in detecting rumor. Our model exhibited a stronger simulation ability, better scalability, and better persuasive ability.

\section{The LOSIRD Model}

\subsection{Problem Statement}
We formulated this rumor detection task as a hybrid task that combines the evidence retrieval sub-task and rumor prediction sub-task.

The evidence retrieval sub-task was defined as: Given a claim, the target of this sub-task was to match textural evidence from Wikipedia and reason the relationship between those potential evidence sentences and the given claim as “SUPPORTED”, “REFUTED” or “NOT ENOUGH INFO (NEI)”. We defined the Wikipedia as an objective information corpus: $ Wiki=\{D_1,D_2,...,D_{|w|}\} $, $ D_i $ as a document from Wikipedia. One document comprised several sentences describing one entity in Wikipedia. The goal of this sub-task was to retrieve evidence, classify the relationship of the evidence and the given claim $C$ i.e., $f_{ERM}:C\rightarrow \{(y_e, E);E\in Wiki\}$, $y_e$ is the predicted evidence label of the claim, $E$ is retrieved evidence set of the claim which contains several sentence-level pieces of evidence.

The rumor prediction sub-task was defined as: Given a claim, that claim’s replies, that claim’s retrieved evidence set and evidence label, the model detected whether the claim was a rumor or non-rumor and provide the evidence. We defined the rumor dataset in this sub-task as $\Psi = \{T_1,T_2,...,T_{|\Psi|}\}$, where $T_i$ is a tweet in the dataset. $T_{i}=\{C_i,P_i,E_i,y_{e_{i}}\}$, where $C_i$ is the ith source tweet in the rumor dataset, $P_i$ is the  source tweet’s reply posts, $E_i$ and $y_{e_i}$ are the corresponding $C_i$ retrieved evidence set and evidence label. Given a tweet $T$ the function of this task was defined as $f_{RDM}:T\rightarrow \{(y_r,E);E\in Wiki\}$, $y_r$ was the predicted rumor label.

% 图4
\begin{figure*}[t]
	\centering
	\includegraphics[width=0.95\linewidth]{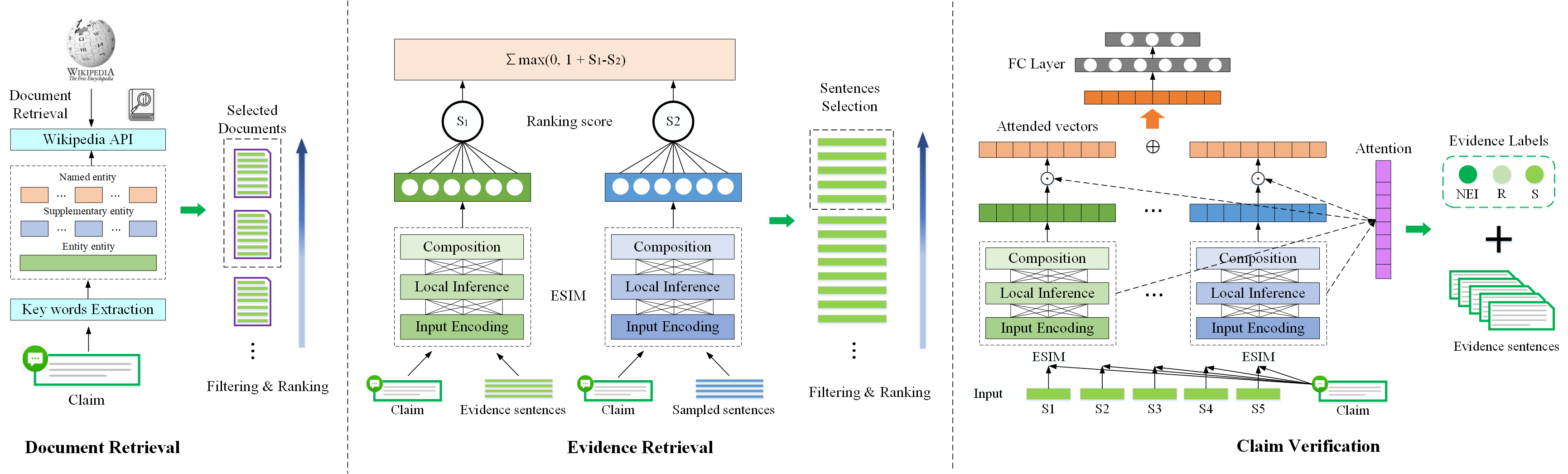}
	\caption{The architecture of the ERM.}
	\label{fig:f4}
\end{figure*}

\subsection{ERM}
Mainly following \citep{liu2020fine} and \citep{hanselowski2018ukp}, we adopted a three-step pipeline module for retrieval evidence, called ERM. The architecture of the ERM is shown in Fig. 3. It contains three main steps ,i.e., Document retrieval: employs a keyword matching algorithm to crawling related files in Wikipedia. Evidence retrieval: extract sentence-level evidence from the retrieved articles. Claim verification: based on the sentence-level evidence, it predicts the relationship of the claim and the evidence as “Supported”, “Refuted”, or “NEI”. Specifically, the ERM first leverages semantic NLP tool-kits to extract potential entities from the given a claim. With the parsed entities, top k highest-ranked Wikipedia articles were filtered by the MediaWiki API. And then, from those retrieved documents, the ERM extracts objective facts as the predicted evidence in the form of sentences that are relevant for the claim. Finally, a verification component of ERM performs prediction over the given statement and retrieved evidence, and verifies the relationship between the claim and evidence as supporting, refuting, or NEI. 

\subsection{RDM}
Fig. 4 shows the structure of RDM. Since the source tweet forms different topology between replies and the evidence, two heterogeneous graph objects, the conversation tree-shaped graph and the evidence star-shaped graph, were structured in RDM.
% 图5
\begin{figure*}[h]
	\centering
	\includegraphics[width=0.90\linewidth]{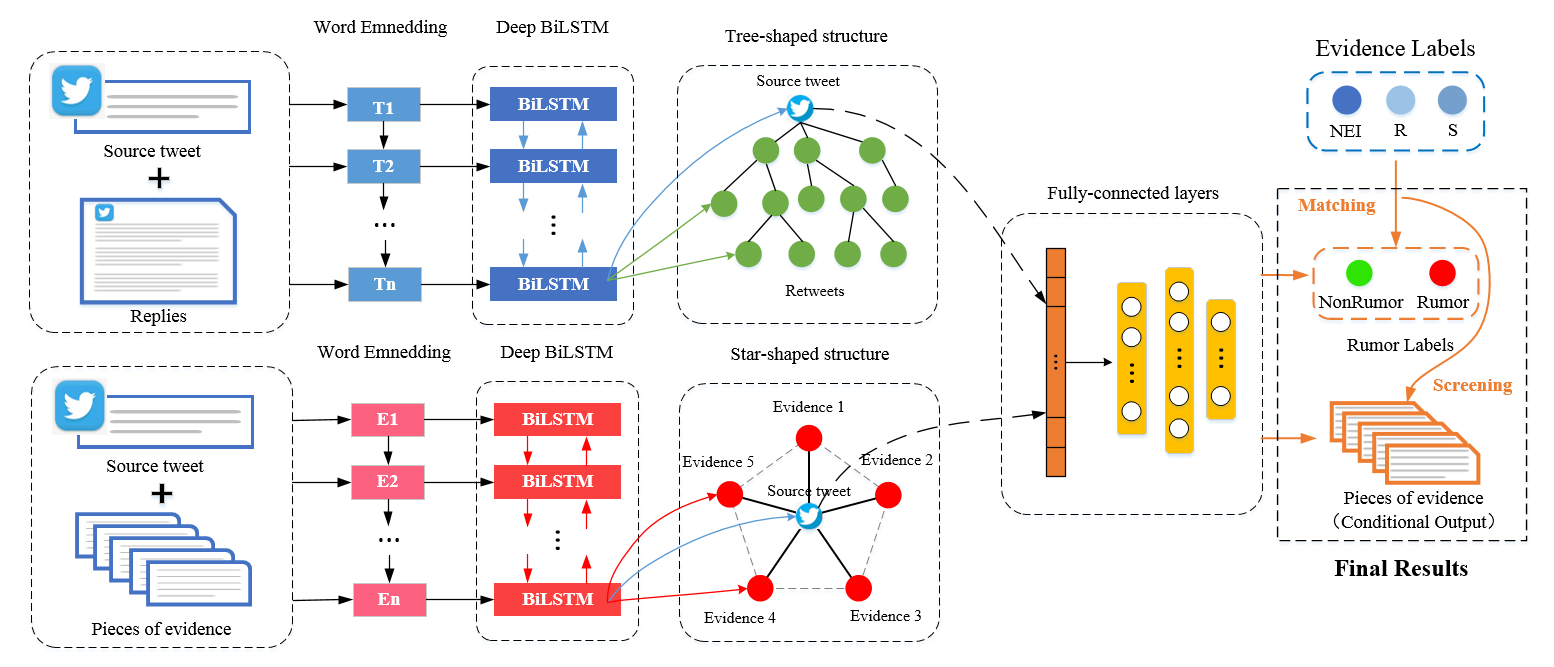}
	\caption{The architecture of the RDM.}
	\label{fig:f5}
\end{figure*}
\subsubsection{Two heterogeneous graphs}
\textbf{Conversation tree-shaped structure} is a peculiar reply relationship topology that forms naturally from social media and carries a vital clue for rumor detection \citep{belkaroui2014conversation,pace2016structure}. Of note, the conversation structure is tree-shaped. Where the root of the tree is the source tweet, each node represents a comment, and each node is connected by its reply relationship.

\textbf{Evidence star-shaped structure} suggests that each piece of evidence is a supplementary description to the source tweet, hence each evidence sentence directly related to the source tweet forms a star topology. In this star-shaped structure, the node of the source tweet is in the center and all the evidence nodes surround the source tweet representing an angle in the star structure.
\subsubsection{Rumor Detection Module}
The rumor detection module contains four components: (1) a word vector encoding component. (2) a sentence embedding component. (3) a graph processing component. (4) a classifier component.

In RDM, a deep BiLSTM was utilized to extract the information among words and generate a sentence representation. The obtained sentence vector was passed into a graph processing component, the GraphSAGE \citep{hamilton2017inductive} model was used as its backbone. The GraphSAGE effectively handled variable graphs. Since the output of the previous component is a set of sentence vectors does not contain structural information. Therefore, before passing this information into the graph processing component two graph objects, the conversation tree-shaped object and the evidence star-shaped object were constructed respectively.

The creation of the conversation graph object:
\begin{equation}
\begin{split}
&G_p=(V_p,E_p) \\
&V_p=[c,p_1,p_2,..,p_j] \\
&E_p=\{(c,p_1),...(p_n,p_m),...\}
\end{split}
\end{equation}
where $G_p$ is the ith event’s conversation graph object, it’s vertex set is $V_p$ and edge set is $E_p$. The vertex set includes all the post in the event, and the edge set $E_p$ means the reply relationship between each post. $c$ and $p_j$ are the tweet embedding results from the BiLSTM component, we selected the last hidden state of BiLSTM as a sentence embedding result.

The creation of the evidence graph object:
\begin{equation}\label{key}
\begin{split}
&G_e=(V_e,E_e)\\
&V_e=[c,e_1,e_2,..,e_k]\\
&E_e=\{(c,e_1),(c,e_2),...,(c,e_k)\}
\end{split}
\end{equation}
where $G_e$ is an evidence graph object consisting of a vertex set $V_e$  and an edge set  $E_e$. The vertex set $V_e$ includes a source post $c$ and evidence sentences, while the edge set $E_e$ represents the relationship between the evidence and the source post, $e_k$ represents the evidence sentence embedding results from the BiLSTM component.

At the beginning of the forward propagation step, the feature of each node was assigned to the nodes in the hidden state as follows:
\begin{equation}\label{key}
\begin{split}
&[h^0_{p_0},h^0_{p_1},...,h^0_{p_j}]\longleftarrow [c,p_1,p_2,..,p_j] \\
&[h^0_{e_0},h^0_{e_1},...,h^0_{e_k}]\longleftarrow [c,e_1,e_2,..,e_k]
\end{split}
\end{equation}
where $h^0_{p_j},h^0_{e_k}$ are the initial hidden states of the nodes of the conversation graph object and the evidence graph object in GraphSAGE.

The node’s hidden state in GraphSAGE updates by constantly aggregating its immediate neighbors’ hidden state, combining them with its own state and generate it’s new hidden state.This process makes the nodes gain incrementally richer information \citep{hamilton2017inductive}:
\begin{equation}\label{key}
\begin{split}
&h^k_{p_{N(v)}}\longleftarrow AGG^{pool}_{k}(\{h^{k-1}_{p_u},\forall u\in N(V_p)\}) \\
&h^k_{p_v}\longleftarrow \sigma (W^k_p\cdot CON(h^{k-1}_{p_v},h^k_{p_{N(v)}}))
\end{split}
\end{equation}
\begin{equation}\label{key}
\begin{split}
&h^k_{e_{N(v)}}\longleftarrow AGG^{pool}_{k}(\{h^{k-1}_{e_u},\forall u\in N(V_e)\}) \\
&h^k_{e_v}\longleftarrow \sigma (W^k_e\cdot CON(h^{k-1}_{e_v},h^k_{e_{N(v)}}))
\end{split}
\end{equation}
where $h^k_{p_{N(v)}},h^k_{e_{N(v)}}$ is the aggregated their neighborhood vectors, $k$ is the depth of the information transmission updates (the number of times the graph information is updated), $N$ is the neighborhood function, $N(v)$ is the set of the node’s immediate neighborhood, and $AGG^{pool}_{k}$ is the aggregation function and $CON$ is the concatenation function.

Three aggregators are provided in GraphSAGE, and in this article we chose the Max Pooling aggregator. Here's the formula:
\begin{equation}\label{key}
\begin{split}
AGG^{pool}_{k}=&max(\{\sigma (W_{pool}\cdot h^k_{graph_{N(v)}}+\\ 
&b_{pool}),\forall u_i\in N(v)\})
\end{split}
\end{equation}
where $max$ is the element-wise max operator, and $\sigma$ is a nonlinear activation function.

After $k$ iterations of information transmission based on the conversation structure and star structure, final representations of the conversation embedding results and evidence embedding results were obtained:
\begin{equation}\label{key}
\begin{split}
&p\longleftarrow h^k_{p_v},\forall v\in V_p \\
&e\longleftarrow h^k_{e_v},\forall v\in V_e
\end{split}
\end{equation}
$p,e$ are the replies and the evidence of the ith event. Max aggregator is used to aggreate the information into fixed size.

Thereafter, the information of these two parts concatenated together then passed into a multilayer perceptron for the final prediction. The formula is as follows:
\begin{equation}\label{key}
y_r=Softmax(V\cdot (p\oplus e)+b_y)
\end{equation}
where $V$ and $b_y$ are parameters in the output layer.

\section{Experiment and Results}
\subsection{Datasets}
Fever dataset was used to train the evidence retrieval module. The statistic of the FEVER dataset is shown in Table 2. 
Two widely used rumor datasets, PHEME 2017 and PHEME 2018\footnote{https://figshare.com/articles/dataset/}, were used to train and evaluate the whole proposed model, as shown in Table 3\footnote{This study have met the terms of accessing and using these rumor datasets.}.
\begin{table}[]
	%\centering
	\caption{The statistic of FEVER dataset.
	}
	\label{tab:my-table}
	\begin{tabular}{|c|c|c|c|}
		\hline
		\textbf{Split}    & \textbf{SUPPORTED} & \textbf{REFUTED} & \textbf{NEI}   \\ \hline
		Train & 80,035     & 29,775   & 35,659 \\ \hline
		Dev      & 6,666      & 6,666    & 6,666  \\ \hline
		Test     & 6,666      & 6,666    & 6,666  \\ \hline
	\end{tabular}
\end{table}

\begin{table}[]
	\caption{The statistic of rumor datasets.
	}
	\centering
	\label{tab:my-table}
	\begin{tabular}{|c|c|c|}
		\hline
		\textbf{Statistic}          & \textbf{PHEME2017} & \textbf{PHEME2018} \\ \hline
		Users                       & 49,345              & 50,593              \\ \hline
		Posts                       & 103,212             & 105,354             \\ \hline
		Events  & 5,802               & 6,425               \\ \hline
		Avg posts/event             & 17.8                & 16.3                \\ \hline
		Rumor                       & 1,972                & 2,402                \\ \hline
		Non-rumor                   & 3,830                & 4,023                \\ \hline
	\end{tabular}
\end{table}

\subsection{Experimental Setup}
To evaluate the rumor detection performance of our model, we compared our proposed models with other popular rumor detection models, including some of the current state-of-the-art models. In the text processing stage, we clean the text information by removing useless expressions and symbols, uniform case, etc. We use Twitter 27B pre-trained GloVe data with 200 dimensions for word embedding and set the maximum vocabulary to 80,000. For the rumor detection module The hidden size of BiLSTM is 128, and the number of layers is 2. The batch size of graphSAGE is 64. We use Adam with a 0.0015 learning rate to optimize the model, with the dropout rate set to 0.5. For the evidence retrieval, we set the learning rate in ESIM is 0.002, drop out rate is 0, batch size is 64, activation fuction is relu. For the claim verification, we set the the learning rate in ESIM is 0.002, drop out rate is 0.1, batch size is 128,activation fuction is relu.We split the datasets, reserve 10\% of the events as the validation set, and the rest in a ratio of 3:1 for training and testing partitions. 

\begin{itemize}
	\item \textbf{CNN}: a convolutional neural model for rumor detection \citep{chen2017ikm}.
	\item \textbf{BiLSTM}: a bidirectional LSTM model for debunking rumors \citep{augenstein2016stance}.
	\item \textbf{BERT}: a fine-tuned BERT to detect rumors \citep{devlin2019bert}.
	\item \textbf{CSI}: a state-of-the-art model detecting rumor by scoring users based on their behavior \citep{ruchansky2017csi}.
	\item \textbf{DEFEND}: a state-of-the-art model learns the correlation between the source article's sentences and user profiles \citep{shu2019defend}.
	\item \textbf{RDM}: a state-of-the-art model integrating GRU and reinforcement learning to detect rumors at an early stage \citep{zhou2019early}.
	\item \textbf{CSRD}: a state-of-the-art model that detect rumors by modeling conversation structure \citep{li2020exploiting}.
	\item \textbf{LOSIDR}: our model, leverages objective facts and subjective views for interpretable rumor detection. 
\end{itemize}
% 表二
\begin{table*}[t]
	\centering
	\caption{\label{t3}
		Main Experimental results. The best model and the best competitor are highlighted by bold and underline.
	}
	\scalebox{1}{
		\setlength{\tabcolsep}{3mm}{
			\begin{tabular}{|l|c|c|c|c|c|c|c|c|}
				\hline
				\multirow{2}{*}{\textbf{Method}}& \multicolumn{4}{c|} {\textbf{PHEME 2017}} & \multicolumn{4}{c|}{\textbf{PHEME 2018}} \\
				\cline{2-9}
				&\textbf{Acc}  &\textbf{Pre}  &\textbf{Rec}  &\textbf{F1}  &\textbf{Acc}  &\textbf{Pre}  &\textbf{Rec} &\textbf{F1}  \\
				\hline
				
				CNN&0.787  &0.737  &0.702  &0.710  &0.795  &0.731  &0.673  &0.686  \\
				
				BiLSTM&0.795  &0.763  &0.691  &0.725  &0.794  &0.727  &0.677  &0.701  \\
				
				BERT&0.865  &0.859  &0.851  &0.855  &0.844  &0.834  &0.835  &0.835  \\
				
				CSI&0.857  &0.843  &0.859  &0.851  &0.851  &0.836  &0.855  &0.845  \\

				DEFEND&0.868  &0.867  &0.859  &0.863  &0.863  &0.857  &0.859  &0.858  \\
				
				RDM&0.873  &0.817  &0.823  &0.820  &0.858  &0.847  &0.859  &0.852  \\
				
				CSRD &0.900& 0.893& 0.869& 0.881 &  0.919& 0.892& 0.923& 0.907 \\
				\hline
				LOSIRD&\textbf{0.914}  &\textbf{0.915}  &\textbf{0.900} &\textbf{0.906}  &\textbf{0.925}  &\textbf{0.922}  &\textbf{0.924}  &\textbf{0.923}  \\
				\hline
	\end{tabular}}}
\end{table*}

\subsection{Experimental Results}
The main experimental results are shown in Table 4. The LOSIRD outperformed the other best-competing methods on PHEME 17 and PHEME 18. Its accuracy was 91.4\% in PHEME 2017 and 92.5\% in PHEME 2018. Moreover, the precision, recall, and F1 were all higher than 90\% in both two datasets. Such promising results confirmed the effectiveness of evidence information and the topology message processing method in rumor detection. For the CNN, BiLSTM, DEFEND, and RDM models, they typically concatenated posts as a single line based on the publish time, while ignoring the conversation structure information. Nonetheless, the structure was crucial for encoding the posts to comprehensive and precise representations. The CSI and CRNN processed topology information, but only the subjective information was adopted in those models causing insufficiency in information extraction.

\subsection{Evidence Impact Study}
In this section, we discussed whether the evidence facilitates rumor detection and determined the extent of the impact of the evidence in debunking rumor. Notably, the evaluated datasets were the PHEME 2017 and PHEME 2018. 
\subsubsection{Distribution of Retrieved Evidence}
To accurately evaluate the retrieved evidence, the distribution of the retrieved evidence based on its evidence label was analyzed.
Two pie charts were constructed to reflect their distribution situations. As shown in Fig. 5, most of the retrieved evidence was irrelevant to the given claim, and about 14.8\% of the retrieved evidence sentences had sufficient information that supports or refutes the given claim. Despite the proportion of supports and refutes being not large, this result was commendable and better than our expectations. 
% 图7
\begin{figure}[t]
	\centering
	\includegraphics[width=1\linewidth]{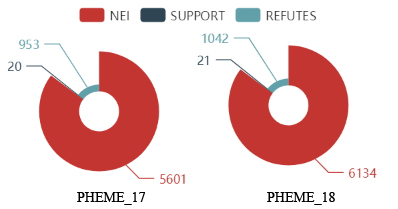}
	\caption{The distribution of the retrieved evidence.}
	\label{f7}
\end{figure}

\subsubsection{Retrieved Evidence Probability Analysis}
We further evaluated the impact of evidence by statistically calculating the probability gap between rumor in original data and rumor in data that labeled refutes. The outcome is shown in Table 5. The probabilities of rumor in original data were about 35\% in both datasets, while the probabilities of rumor in data that labeled refutes were around 73\%, which was much higher than in original data. Specifically, rumor in data that labeled refutes increased to 42.5\% on PHEME 17 and 34.3\% on PHEME 18. This strongly confirmed that the retrieved evidence was a vital clue for rumor detection.
\begin{table}[]
	\centering
	\caption{Retrieved evidence probability analysis result.
	}
	\label{tab:my-table}
	\begin{tabular}{|c|m{1.35cm}|m{1.4cm}|m{1.6cm}|}
		\hline
		\textbf{Dataset}    & \textbf{Original} & \textbf{Refutes} & \textbf{Increment}   \\ \hline
		PHEME 17 & 33.30\%     & 75.80\%   & 42.50\% \\ \hline
		PHEME 18     & 36.60\%       & 70.92\%    & 34.30\%  \\ \hline
	\end{tabular}
\end{table}

\subsubsection{Influence Analysis of the Evidence on Deep Learning Model}
% 图8
\begin{figure*}[]
	\centering
	\includegraphics[width=0.93\linewidth]{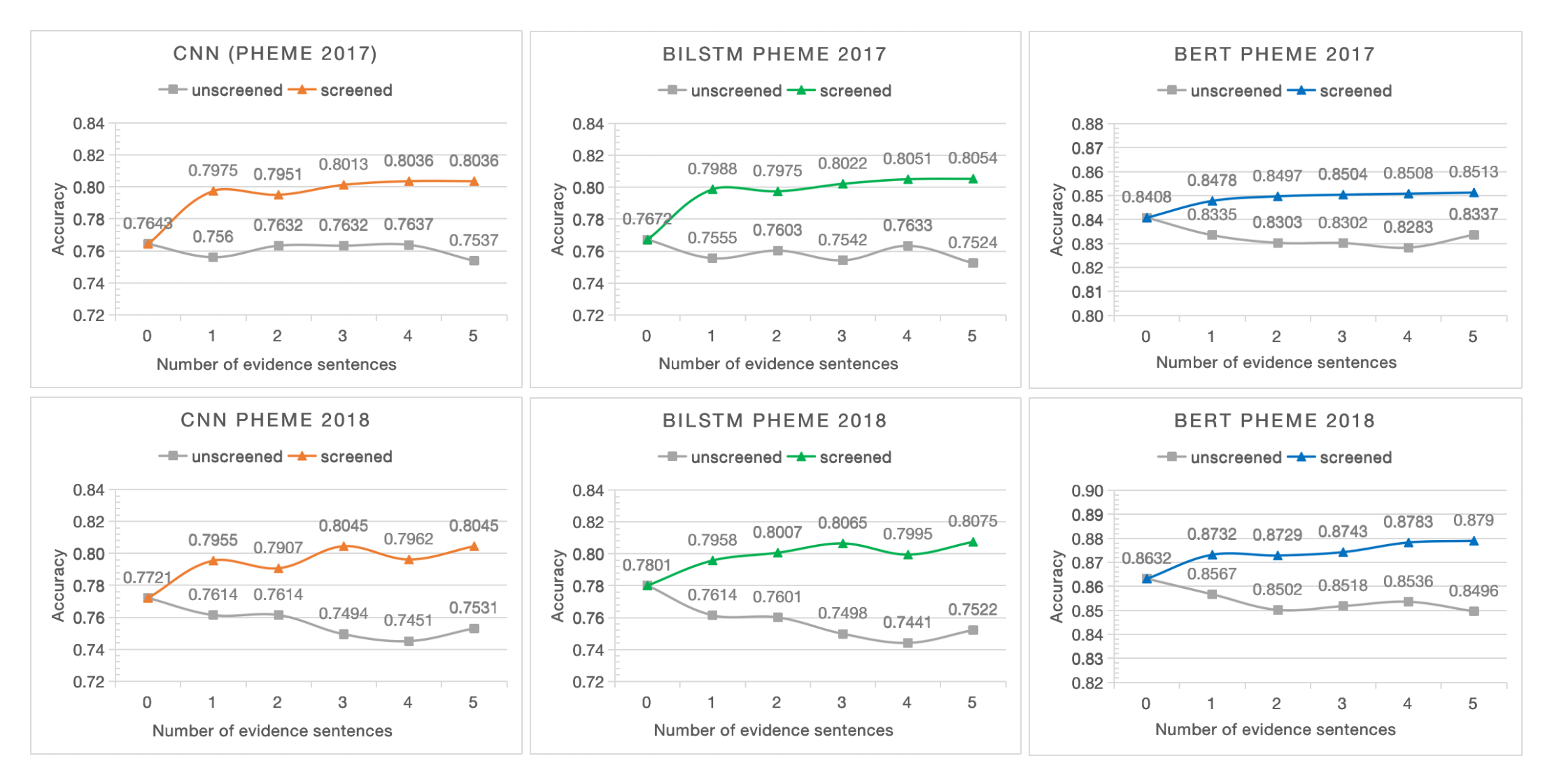}
	\caption{Influence analysis of the evidence on the performance of deep learning models.}
	\label{f8}
\end{figure*}
To further illustrate the influence of evidence on rumor detection and analyze the impact of evidence on deep learning models, three NLP models, CNN, BILSTM, and BERT, were deliberately selected as the examination models in this subsection. We concatenated the suspicious claim and its evidence sentences and inputted them into the three models, respectively. The experimental results shows in Fig. 6. The horizontal axis represented a different number of evidence sentences, 0 means only source tweet, while 1 to 5 means source tweet plus 1 to 5 evidence sentences. Also, this paper analyzed the performance before and after the evidence was filtered which was represented as each chart with two lines i.e., one for the unscreened evidence (filter the NEI evidence) and the other for the screened evidence. The broken lines of unscreened in all the charts showed a downward trend. This indicated that the NEI evidence contained a certain amount of useless information there by making the detection process harder. Furthermore, after dropping the NEI evidence, all the models achieved an improvement by an increase of 5\% accuracy on average. This demonstrated that the filtered evidence significantly helps the deep learning models in debunking rumors.

\subsection{Early Detection Performance}
To evaluate the early rumor detection performance of our model, 9 test sets that reflected real-world scenarios of rumors spreading on Twitter were created. Each test set included a different number of replies, ranging between 5 replies and 45 replies. The test subset was sampled based on the publication timestamp. As shown in Fig. 7, even though the number of posts was only 5, our LOSIRD model had more than 91\% accuracy in PHEME 2017 dataset and PHEME 2018 dataset. Additionally, the broken line diagram showed that the curve of our model was significantly stable, indicating satisfactory robustness and high performance in early rumor detection. Besides, our model effectively made use of the objective information from Wikipedia, hence, it did not rely on subjective information from replies of the users thereby achieving satisfactory performance in the early stage of rumor propagation.
% 图12
\begin{figure}[]
	\centering
	\includegraphics[width=1\linewidth]{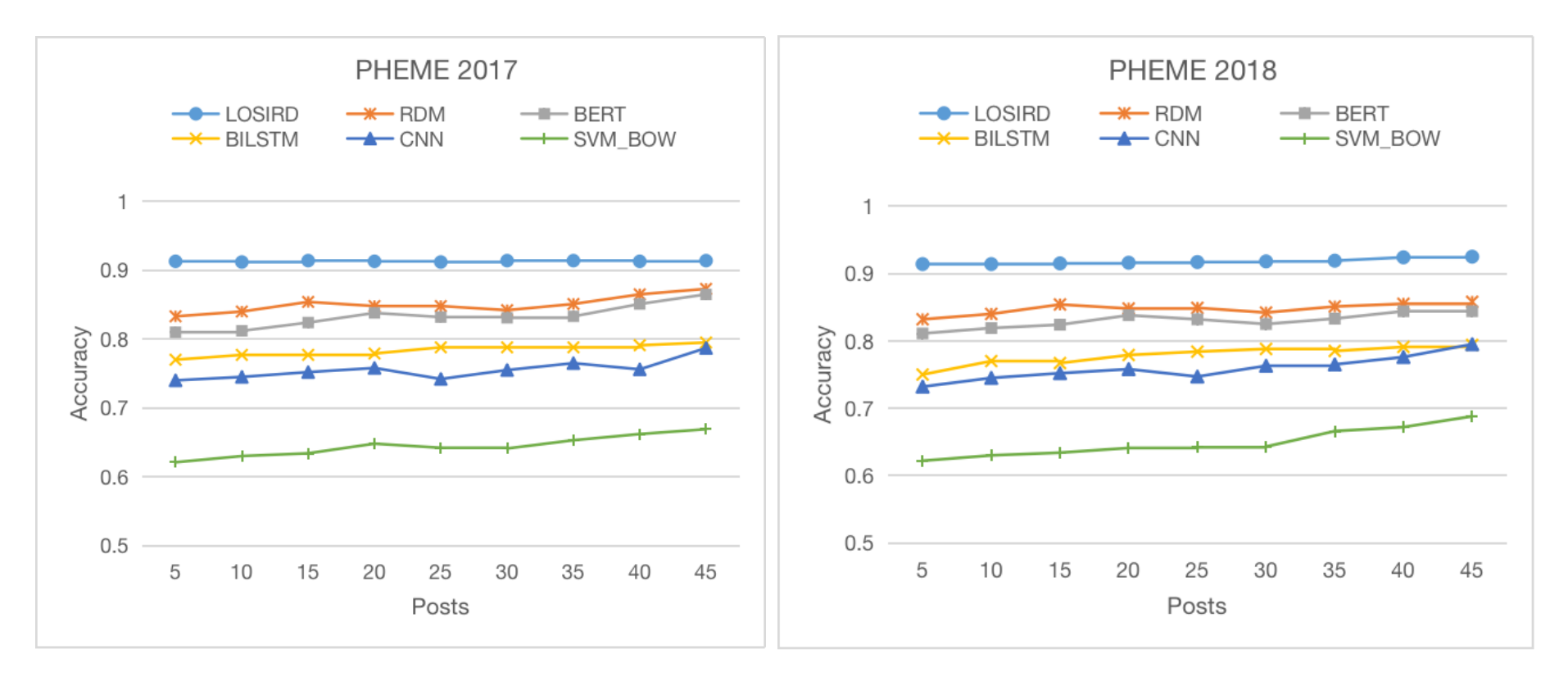}
	\caption{Early rumor detection performance.}
	\label{f12}
\end{figure}

\section{Conclusion}
In this paper, we proposed a LOSIRD, a novel interpretative model for rumor detection. Notably, the LOSIRD debunking rumor mechanism depends on both objective facts and subjective views. Objective fact sentences retrieved from 5,416,537 Wikipedia articles were sufficiently utilized to help LOSIRD in analyzing the veracity of a suspicious claim. Meanwhile, the information in subjective views was extracted by simulating the propagation of subjective views based on the conversation structure. Results on two public Twitter datasets showed that our model improved rumor detection performance by a certain margin compared to the state-of-the-art baselines. Further, we analyzed the impact of objective facts for rumor detection and analyzed the effectiveness of the conversation structure. The experiments revealed that both objective facts and subjective views were vital clues for debunking rumor. Moreover, we believe that our model will be used for rumor detection and other text classification tasks on social media.

\section*{Acknowledgments}
This work was funded in part by Qualcomm through a Taiwan University Research Collaboration Project and in part by the Ministry of Science and Technology, Taiwan, under grant MOST 109-2221-E-006-173 and NCKU B109-K027D.

\bibliographystyle{acl_natbib}
\bibliography{anthology,acl2021}

\begin{thebibliography}{35}
\expandafter\ifx\csname natexlab\endcsname\relax\def\natexlab#1{#1}\fi

\bibitem[{Augenstein et~al.(2016)Augenstein, Rockt{\"a}schel, Vlachos, and
  Bontcheva}]{augenstein2016stance}
Isabelle Augenstein, Tim Rockt{\"a}schel, Andreas Vlachos, and Kalina
  Bontcheva. 2016.
\newblock \href {https://doi.org/10.18653/v1/D16-1084} {Stance detection with
  bidirectional conditional encoding}.
\newblock In \emph{Proceedings of the 2016 Conference on Empirical Methods in
  Natural Language Processing}, pages 876--885, Austin, Texas. Association for
  Computational Linguistics.

\bibitem[{Belkaroui et~al.(2014)Belkaroui, Faiz, and
  Elkhlifi}]{belkaroui2014conversation}
Rami Belkaroui, Rim Faiz, and Aymen Elkhlifi. 2014.
\newblock Conversation analysis on social networking sites.
\newblock In \emph{2014 Tenth International Conference on Signal-Image
  Technology and Internet-Based Systems}, pages 172--178. IEEE.

\bibitem[{Castillo et~al.(2011)Castillo, Mendoza, and
  Poblete}]{castillo2011information}
Carlos Castillo, Marcelo Mendoza, and Barbara Poblete. 2011.
\newblock \href {https://doi.org/10.1145/1963405.1963500} {Information
  credibility on twitter}.
\newblock In \emph{Proceedings of the 20th International Conference on World
  Wide Web, {WWW} 2011, Hyderabad, India, March 28 - April 1, 2011}, pages
  675--684. {ACM}.

\bibitem[{Chen et~al.(2018)Chen, Li, Yin, and Zhang}]{chen2018call}
Tong Chen, Xue Li, Hongzhi Yin, and Jun Zhang. 2018.
\newblock Call attention to rumors: Deep attention based recurrent neural
  networks for early rumor detection.
\newblock In \emph{Pacific-Asia conference on knowledge discovery and data
  mining}, pages 40--52. Springer.

\bibitem[{Chen et~al.(2017)Chen, Liu, and Kao}]{chen2017ikm}
Yi-Chin Chen, Zhao-Yang Liu, and Hung-Yu Kao. 2017.
\newblock \href {https://doi.org/10.18653/v1/S17-2081} {{IKM} at
  {S}em{E}val-2017 task 8: Convolutional neural networks for stance detection
  and rumor verification}.
\newblock In \emph{Proceedings of the 11th International Workshop on Semantic
  Evaluation ({S}em{E}val-2017)}, pages 465--469, Vancouver, Canada.
  Association for Computational Linguistics.

\bibitem[{Devlin et~al.(2019)Devlin, Chang, Lee, and
  Toutanova}]{devlin2019bert}
Jacob Devlin, Ming-Wei Chang, Kenton Lee, and Kristina Toutanova. 2019.
\newblock \href {https://doi.org/10.18653/v1/N19-1423} {{BERT}: Pre-training of
  deep bidirectional transformers for language understanding}.
\newblock In \emph{Proceedings of the 2019 Conference of the North {A}merican
  Chapter of the Association for Computational Linguistics: Human Language
  Technologies, Volume 1 (Long and Short Papers)}, pages 4171--4186,
  Minneapolis, Minnesota. Association for Computational Linguistics.

\bibitem[{Hamilton et~al.(2017)Hamilton, Ying, and
  Leskovec}]{hamilton2017inductive}
William~L. Hamilton, Zhitao Ying, and Jure Leskovec. 2017.
\newblock \href
  {https://proceedings.neurips.cc/paper/2017/hash/5dd9db5e033da9c6fb5ba83c7a7ebea9-Abstract.html}
  {Inductive representation learning on large graphs}.
\newblock In \emph{Advances in Neural Information Processing Systems 30: Annual
  Conference on Neural Information Processing Systems 2017, December 4-9, 2017,
  Long Beach, CA, {USA}}, pages 1024--1034.

\bibitem[{Hanselowski et~al.(2018)Hanselowski, Zhang, Li, Sorokin, Schiller,
  Schulz, and Gurevych}]{hanselowski2018ukp}
Andreas Hanselowski, Hao Zhang, Zile Li, Daniil Sorokin, Benjamin Schiller,
  Claudia Schulz, and Iryna Gurevych. 2018.
\newblock \href {https://doi.org/10.18653/v1/W18-5516} {{UKP}-athene:
  Multi-sentence textual entailment for claim verification}.
\newblock In \emph{Proceedings of the First Workshop on Fact Extraction and
  {VER}ification ({FEVER})}, pages 103--108, Brussels, Belgium. Association for
  Computational Linguistics.

\bibitem[{Kochkina et~al.(2018)Kochkina, Liakata, and
  Zubiaga}]{kochkina2018all}
Elena Kochkina, Maria Liakata, and Arkaitz Zubiaga. 2018.
\newblock \href {https://www.aclweb.org/anthology/C18-1288} {All-in-one:
  Multi-task learning for rumour verification}.
\newblock In \emph{Proceedings of the 27th International Conference on
  Computational Linguistics}, pages 3402--3413, Santa Fe, New Mexico, USA.
  Association for Computational Linguistics.

\bibitem[{Li et~al.(2020{\natexlab{a}})Li, Ni, and Kao}]{li2020birds}
Jiawen Li, Shiwen Ni, and Hung-Yu Kao. 2020{\natexlab{a}}.
\newblock Birds of a feather rumor together? exploring homogeneity and
  conversation structure in social media for rumor detection.
\newblock \emph{IEEE Access}, 8:212865--212875.

\bibitem[{Li et~al.(2020{\natexlab{b}})Li, Sujana, and Kao}]{li2020exploiting}
Jiawen Li, Yudianto Sujana, and Hung-Yu Kao. 2020{\natexlab{b}}.
\newblock \href {https://doi.org/10.18653/v1/2020.coling-main.473} {Exploiting
  microblog conversation structures to detect rumors}.
\newblock In \emph{Proceedings of the 28th International Conference on
  Computational Linguistics}, pages 5420--5429, Barcelona, Spain (Online).
  International Committee on Computational Linguistics.

\bibitem[{Li et~al.(2019)Li, Zhang, and Si}]{li2019rumor}
Quanzhi Li, Qiong Zhang, and Luo Si. 2019.
\newblock \href {https://doi.org/10.18653/v1/P19-1113} {Rumor detection by
  exploiting user credibility information, attention and multi-task learning}.
\newblock In \emph{Proceedings of the 57th Annual Meeting of the Association
  for Computational Linguistics}, pages 1173--1179, Florence, Italy.
  Association for Computational Linguistics.

\bibitem[{Liu et~al.(2015)Liu, Nourbakhsh, Li, Fang, and Shah}]{liu2015real}
Xiaomo Liu, Armineh Nourbakhsh, Quanzhi Li, Rui Fang, and Sameena Shah. 2015.
\newblock \href {https://doi.org/10.1145/2806416.2806651} {Real-time rumor
  debunking on twitter}.
\newblock In \emph{Proceedings of the 24th {ACM} International Conference on
  Information and Knowledge Management, {CIKM} 2015, Melbourne, VIC, Australia,
  October 19 - 23, 2015}, pages 1867--1870. {ACM}.

\bibitem[{Liu and Wu(2018)}]{liu2018early}
Yang Liu and Yi{-}fang~Brook Wu. 2018.
\newblock \href
  {https://www.aaai.org/ocs/index.php/AAAI/AAAI18/paper/view/16826} {Early
  detection of fake news on social media through propagation path
  classification with recurrent and convolutional networks}.
\newblock In \emph{Proceedings of the Thirty-Second {AAAI} Conference on
  Artificial Intelligence, (AAAI-18), the 30th innovative Applications of
  Artificial Intelligence (IAAI-18), and the 8th {AAAI} Symposium on
  Educational Advances in Artificial Intelligence (EAAI-18), New Orleans,
  Louisiana, USA, February 2-7, 2018}, pages 354--361. {AAAI} Press.

\bibitem[{Liu et~al.(2020)Liu, Xiong, Sun, and Liu}]{liu2020fine}
Zhenghao Liu, Chenyan Xiong, Maosong Sun, and Zhiyuan Liu. 2020.
\newblock \href {https://doi.org/10.18653/v1/2020.acl-main.655} {Fine-grained
  fact verification with kernel graph attention network}.
\newblock In \emph{Proceedings of the 58th Annual Meeting of the Association
  for Computational Linguistics}, pages 7342--7351, Online. Association for
  Computational Linguistics.

\bibitem[{Lu and Li(2020)}]{lu2020gcan}
Yi-Ju Lu and Cheng-Te Li. 2020.
\newblock \href {https://doi.org/10.18653/v1/2020.acl-main.48} {{GCAN}:
  Graph-aware co-attention networks for explainable fake news detection on
  social media}.
\newblock In \emph{Proceedings of the 58th Annual Meeting of the Association
  for Computational Linguistics}, pages 505--514, Online. Association for
  Computational Linguistics.

\bibitem[{Ma et~al.(2016)Ma, Gao, Mitra, Kwon, Jansen, Wong, and
  Cha}]{ma2016detecting}
Jing Ma, Wei Gao, Prasenjit Mitra, Sejeong Kwon, Bernard~J. Jansen, Kam{-}Fai
  Wong, and Meeyoung Cha. 2016.
\newblock \href {http://www.ijcai.org/Abstract/16/537} {Detecting rumors from
  microblogs with recurrent neural networks}.
\newblock In \emph{Proceedings of the Twenty-Fifth International Joint
  Conference on Artificial Intelligence, {IJCAI} 2016, New York, NY, USA, 9-15
  July 2016}, pages 3818--3824. {IJCAI/AAAI} Press.

\bibitem[{Ma et~al.(2017)Ma, Gao, and Wong}]{ma2017detect}
Jing Ma, Wei Gao, and Kam-Fai Wong. 2017.
\newblock \href {https://doi.org/10.18653/v1/P17-1066} {Detect rumors in
  microblog posts using propagation structure via kernel learning}.
\newblock In \emph{Proceedings of the 55th Annual Meeting of the Association
  for Computational Linguistics (Volume 1: Long Papers)}, pages 708--717,
  Vancouver, Canada. Association for Computational Linguistics.

\bibitem[{Ma et~al.(2018)Ma, Gao, and Wong}]{ma2018rumor}
Jing Ma, Wei Gao, and Kam-Fai Wong. 2018.
\newblock \href {https://doi.org/10.18653/v1/P18-1184} {Rumor detection on
  {T}witter with tree-structured recursive neural networks}.
\newblock In \emph{Proceedings of the 56th Annual Meeting of the Association
  for Computational Linguistics (Volume 1: Long Papers)}, pages 1980--1989,
  Melbourne, Australia. Association for Computational Linguistics.

\bibitem[{Malon(2018)}]{malon2018team}
Christopher Malon. 2018.
\newblock \href {https://doi.org/10.18653/v1/W18-5517} {Team papelo:
  Transformer networks at {FEVER}}.
\newblock In \emph{Proceedings of the First Workshop on Fact Extraction and
  {VER}ification ({FEVER})}, pages 109--113, Brussels, Belgium. Association for
  Computational Linguistics.

\bibitem[{Merigo et~al.(2016)Merigo, Palacios-Marques, and
  Zeng}]{merigo2016subjective}
Jose~M Merigo, Daniel Palacios-Marques, and Shouzhen Zeng. 2016.
\newblock Subjective and objective information in linguistic multi-criteria
  group decision making.
\newblock \emph{European Journal of Operational Research}, 248(2):522--531.

\bibitem[{Monti et~al.(2019)Monti, Frasca, Eynard, Mannion, and
  Bronstein}]{monti2019fake}
Federico Monti, Fabrizio Frasca, Davide Eynard, Damon Mannion, and Michael~M
  Bronstein. 2019.
\newblock Fake news detection on social media using geometric deep learning.

\bibitem[{Nguyen(2019)}]{nguyen2019graph}
Thanh~Tam Nguyen. 2019.
\newblock Graph-based rumour detection for social media.
\newblock Technical report.

\bibitem[{Pace et~al.(2016)Pace, Buzzanca, and Fratocchi}]{pace2016structure}
Stefano Pace, Stefano Buzzanca, and Luciano Fratocchi. 2016.
\newblock The structure of conversations on social networks: Between dialogic
  and dialectic threads.
\newblock \emph{International Journal of Information Management},
  36(6):1144--1151.

\bibitem[{Rath et~al.(2017)Rath, Gao, Ma, and Srivastava}]{rath2017retweet}
Bhavtosh Rath, Wei Gao, Jing Ma, and Jaideep Srivastava. 2017.
\newblock \href {https://doi.org/10.1145/3110025.3110121} {From retweet to
  believability: Utilizing trust to identify rumor spreaders on twitter}.
\newblock In \emph{Proceedings of the 2017 {IEEE/ACM} International Conference
  on Advances in Social Networks Analysis and Mining 2017, Sydney, Australia,
  July 31 - August 03, 2017}, pages 179--186. {ACM}.

\bibitem[{Ruchansky et~al.(2017)Ruchansky, Seo, and Liu}]{ruchansky2017csi}
Natali Ruchansky, Sungyong Seo, and Yan Liu. 2017.
\newblock \href {https://doi.org/10.1145/3132847.3132877} {{CSI:} {A} hybrid
  deep model for fake news detection}.
\newblock In \emph{Proceedings of the 2017 {ACM} on Conference on Information
  and Knowledge Management, {CIKM} 2017, Singapore, November 06 - 10, 2017},
  pages 797--806. {ACM}.

\bibitem[{Shu et~al.(2019)Shu, Cui, Wang, Lee, and Liu}]{shu2019defend}
Kai Shu, Limeng Cui, Suhang Wang, Dongwon Lee, and Huan Liu. 2019.
\newblock \href {https://doi.org/10.1145/3292500.3330935} {defend: Explainable
  fake news detection}.
\newblock In \emph{Proceedings of the 25th {ACM} {SIGKDD} International
  Conference on Knowledge Discovery {\&} Data Mining, {KDD} 2019, Anchorage,
  AK, USA, August 4-8, 2019}, pages 395--405. {ACM}.

\bibitem[{Sujana et~al.(2020)Sujana, Li, and Kao}]{sujana2020rumor}
Yudianto Sujana, Jiawen Li, and Hung-Yu Kao. 2020.
\newblock \href {https://www.aclweb.org/anthology/2020.aacl-main.3} {Rumor
  detection on {T}witter using multiloss hierarchical {B}i{LSTM} with an
  attenuation factor}.
\newblock In \emph{Proceedings of the 1st Conference of the Asia-Pacific
  Chapter of the Association for Computational Linguistics and the 10th
  International Joint Conference on Natural Language Processing}, pages 18--26,
  Suzhou, China. Association for Computational Linguistics.

\bibitem[{Yang et~al.(2018)Yang, Zheng, Zhang, Cui, Li, and Yu}]{yang2018ti}
Yang Yang, Lei Zheng, Jiawei Zhang, Qingcai Cui, Zhoujun Li, and Philip~S Yu.
  2018.
\newblock Ti-cnn: Convolutional neural networks for fake news detection.
\newblock \emph{arXiv preprint arXiv:1806.00749}.

\bibitem[{Yoneda et~al.(2018)Yoneda, Mitchell, Welbl, Stenetorp, and
  Riedel}]{yoneda2018ucl}
Takuma Yoneda, Jeff Mitchell, Johannes Welbl, Pontus Stenetorp, and Sebastian
  Riedel. 2018.
\newblock \href {https://doi.org/10.18653/v1/W18-5515} {{UCL} machine reading
  group: Four factor framework for fact finding ({H}exa{F})}.
\newblock In \emph{Proceedings of the First Workshop on Fact Extraction and
  {VER}ification ({FEVER})}, pages 97--102, Brussels, Belgium. Association for
  Computational Linguistics.

\bibitem[{Yu et~al.(2017)Yu, Liu, Wu, Wang, and Tan}]{yu2017convolutional}
Feng Yu, Qiang Liu, Shu Wu, Liang Wang, and Tieniu Tan. 2017.
\newblock \href {https://doi.org/10.24963/ijcai.2017/545} {A convolutional
  approach for misinformation identification}.
\newblock In \emph{Proceedings of the Twenty-Sixth International Joint
  Conference on Artificial Intelligence, {IJCAI} 2017, Melbourne, Australia,
  August 19-25, 2017}, pages 3901--3907. ijcai.org.

\bibitem[{Zhong et~al.(2020)Zhong, Xu, Tang, Xu, Duan, Zhou, Wang, and
  Yin}]{zhong2020reasoning}
Wanjun Zhong, Jingjing Xu, Duyu Tang, Zenan Xu, Nan Duan, Ming Zhou, Jiahai
  Wang, and Jian Yin. 2020.
\newblock \href {https://doi.org/10.18653/v1/2020.acl-main.549} {Reasoning over
  semantic-level graph for fact checking}.
\newblock In \emph{Proceedings of the 58th Annual Meeting of the Association
  for Computational Linguistics}, pages 6170--6180, Online. Association for
  Computational Linguistics.

\bibitem[{Zhou et~al.(2019{\natexlab{a}})Zhou, Han, Yang, Liu, Wang, Li, and
  Sun}]{zhou2019gear}
Jie Zhou, Xu~Han, Cheng Yang, Zhiyuan Liu, Lifeng Wang, Changcheng Li, and
  Maosong Sun. 2019{\natexlab{a}}.
\newblock \href {https://doi.org/10.18653/v1/P19-1085} {{GEAR}: Graph-based
  evidence aggregating and reasoning for fact verification}.
\newblock In \emph{Proceedings of the 57th Annual Meeting of the Association
  for Computational Linguistics}, pages 892--901, Florence, Italy. Association
  for Computational Linguistics.

\bibitem[{Zhou et~al.(2019{\natexlab{b}})Zhou, Shu, Li, and
  Lau}]{zhou2019early}
Kaimin Zhou, Chang Shu, Binyang Li, and Jey~Han Lau. 2019{\natexlab{b}}.
\newblock \href {https://doi.org/10.18653/v1/N19-1163} {Early rumour
  detection}.
\newblock In \emph{Proceedings of the 2019 Conference of the North {A}merican
  Chapter of the Association for Computational Linguistics: Human Language
  Technologies, Volume 1 (Long and Short Papers)}, pages 1614--1623,
  Minneapolis, Minnesota. Association for Computational Linguistics.

\bibitem[{Zorio-Grima and Merello(2020)}]{zorio2020consumer}
Ana Zorio-Grima and Paloma Merello. 2020.
\newblock Consumer confidence: Causality links with subjective and objective
  information sources.
\newblock \emph{Technological Forecasting and Social Change}, 150:119760.

\end{thebibliography}

%\appendix

\end{document}